\documentclass[aps,prl,twocolumn,amsmath,amssymb,showkeys]{revtex4-1}
\usepackage{graphicx}
\usepackage{tabularx}
\usepackage{color}

\newcommand{\Ang}{\,\mathrm{\AA}}

\def\K{\,\textrm{K}}
\def\eV{\,\textrm{eV}}
\def\meV{\,\textrm{meV}}

\newcommand{\BMO}{$\mathrm{BiMnO}_3$}
\newcommand{\BFO}{$\mathrm{BiFeO}_3$}
\def\STO{$\mathrm{SrTiO}_3$}
\def\Mn{\textrm{Mn}}

\begin{document}

\title{
Ferroelectricity in $\mathbf{BiMnO_3}$ Thin Films
}
\author{Yun-Peng Wang,$^{1,2} $ J. N. Fry,$^1$  and Hai-Ping Cheng$^{1,2}$ } 
\email[Corr. author: Hai-Ping Cheng,  ]{ hping@ufl.edu }
\affiliation{$^1$Department of Physics, University of Florida,  Gainesville, Florida 32611, USA \\
             $^2$Quantum Theory Project, University of Florida, Gainesville, Florida 32611, USA}


\begin{abstract}

The existence of ferroelectricity in \BMO{} has been a long-standing question for both experimentalists and theorists.
In addition to a highly distorted bulk  structure, the ionic crystal planes cause a large roughness in thin films 
that makes it extremely difficult to nail down the physical mechanisms underlying a possible ferroelectric-ferromagnetic phase.
We approach the problem by including the substrate explicitly to study the polarization. 
With this model, we investigate mono-, di-, and trilayer \BMO{} thin films on \STO{} substrates.
We find that thin film systems have both strong ferromagnetism and strong ferroelectricity.
Substrate constraints weaken the competition between displacements induced by 
stereochemically active Bi-$6s^2$ lone pairs and by Jahn-Teller distortions around Mn ions 
found in the bulk, such that the sum of off-center displacements of Bi ions in bulk \BMO{} nearly cancel.
In \BMO{} thin films, in contrast, all Bi ions displace roughly in parallel, resulting in a strongly polar structure.
We also find spontaneous charge disproportionation of Mn ion pairs in \BMO{} thin films.

\end{abstract}
\maketitle

{\it Introduction.}
Multiferroic materials,  exhibiting magnetic and ferroelectric orders simultaneously, have generated intense interest
due to the rich physics of magnetoelectric coupling and promise in spintronics applications 
\cite{Nature.442.759,Nat.Mater.6.13,Nat.Mater.6.21,Adv.Mater.23.1062}.
Although magnetic and ferroelectric order parameters are predicted to be difficult to manifest simultaneously \cite{JPCB.104.6694}, 
several series of compounds with multiferroic properties were discovered in the last decade 
and have been classified according to the origin of ferroelectric order \cite{Nature.442.759,Nat.Mater.6.13,Nat.Mater.6.21,Adv.Mater.23.1062}.
Most multiferroic materials unfortunately exhibit only weak magnetism, due to canted antiferromagnetic or cycloidal spiral spin order.

Bismuth manganite, \BMO{}, with a highly distorted perovskite structure,
was once regarded as the unique multiferroic compound presenting both ferromagnetic and ferroelectric orders 
Although a C2 space group model was proposed in early studies \cite{JSSC.145.639,PhysRevB.66.064425}, 
 Belik {\it et al.} \cite{JACS.129.971} and Montanari {\it et al.} \cite{PhysRevB.75.220101} 
have argued that bulk \BMO{} fits to A centrosymmetric C2/c space group that forbids ferroelectric order.
Ferromagnetic order in \BMO{} is attributed to the three-dimensional orbital ordering 
\cite{JSSC.145.639,PhysRevB.66.064425,JACS.129.971}, 
and the magnetic structure can be affected by pressure \cite{PhysRevB.78.092404,PhysRevB.82.014401}.
The ferromagnetic Curie temperature is around $100 \K$ \cite{JACS.129.971,PhysRevB.75.220101}.
Each $\Mn^{3+}$ ion carries a magnetic moment of $3.92\,\mu_B$, or close to $4 \,\mu_B$, 
at low temperatures \cite{JACS.129.971}. 
A magneto-electric coupling is manifested by a change in the dielectric constant induced by magnetic fields 
at the Curie temperature \cite{PhysRevB.67.180401}.
Ferroelectric order in bulk \BMO{} was discussed in the literature 
\cite{SolidStateCommun.122.49,JMMM.310.e358,JSSC.145.639,PhysRevB.66.064425,JACS.129.971,PhysRevB.75.220101}.
Inversion symmetry breaking is expected due to the stereochemically active lone pairs $6s^2$ of $\textrm{Bi}^{3+}$,
which make it possible to form cooperative ferroelectric distortions; 
this mechanism is the origin of ferroelectricity in  \BFO{} \cite{PhaseTransitions.79.991}.
Measurements of ferroelectric polarization in bulk \BMO{} were hampered by the high leakage currents 
of polycrystalline samples \cite{PhysRevB.67.180401}, 
and the measured polarization in bulk \BMO{} \cite{SolidStateCommun.122.49,JMMM.310.e358} was attributed to extrinsic factors.
Ferroelectricity in bulk \BMO{}  thus remains in question.

\BMO{} epitaxial thin films have also been investigated for ferroelectricity.
Substrates can stabilize single crystal structure of \BMO{} \cite{ApplPhysRev.2.021304}.
More interestingly, the strain imposed by substrates can change the ferroelectric properties of perovskite materials \cite{ARMR.37.589}.
Again, some \BMO{} thin film samples show ferroelectric properties 
\cite{SolidStateCommun.122.49,APL.84.4971,JAP.109.074104,APL.93.062902,PhysRevB.69.214109,APL.103.062902}, 
while other studies supported an absence of ferroelectricity in \BMO{} thin films 
\cite{J.Korean.Phys.Soc.63.624,Nano.Res.Lett.10.47}

Difficulties in synthesizing high-quality \BMO{} samples make theoretical studies using density functional theory (DFT) indispensable.
Ferroelectric instability in \BMO{} was investigated based on local spin density approximation (LSDA) 
calculations  \cite{PhysRevB.59.8759}.
The stability of the centrosymmetric C2/c phase over the C2 phase was confirmed using LSDA 
with an on-site Hubbard Coulomb correction (LSDA+$U$) \cite{JACS.129.9854} 
and \BMO{} was predicted to be nonpolar under compressive or tensile strain \cite {EPJB.71.435}.
Recent studies showed that LSDA+$U$ and GGA+$U$ (GGA: generalized gradient approximation) 
fail to reproduce the structural distortions and energy gap in C2/c \BMO{}.
The strength of Jahn-Teller distortions around $\Mn^{3+}$ ions is heavily underestimated even with the Hubbard Coulomb 
correction on Mn-$d$ orbitals \cite{PhysRevB.82.245101}.
The LSDA+$U$ and GGA+$U$ energy gap,  $ 0.3$--$0.5 \eV$, is also substantially smaller 
than the $0.9 \eV$  found in experiment \cite{PhysRevB.81.144103}.
A recent calculation using the modified Becke-Johnson exchange potential by Tran and Blaha \cite{PhysRevLett.102.226401} 
gives a $0.75 \eV$ energy gap \cite{SSC.243.65}.
Hybrid functionals are capable of reproducing the structural distortion, but overestimate the energy gap \cite{PhysRevB.91.184113}.
Epitaxial phases of \BMO{} were also studied theoretically in \cite{EPJB.71.435,PhysRevB.91.184113}, 
in which substrate effects were simulated using constraints on the in-plane lattice vectors.
The rationale behind this approach is that thin films satisfying the epitaxial relation with respect to the substrates 
are thick enough so that a periodic bulk description is appropriate.
The C2/c phase was found to be the ground state for the in-plane lattice constant between $3.75$ and $4 \Ang$, 
and thus ferroelectric order is forbidden.\cite{PhysRevB.91.184113}

\begin{figure}[t]
\begin{center}
\includegraphics[width=\linewidth]{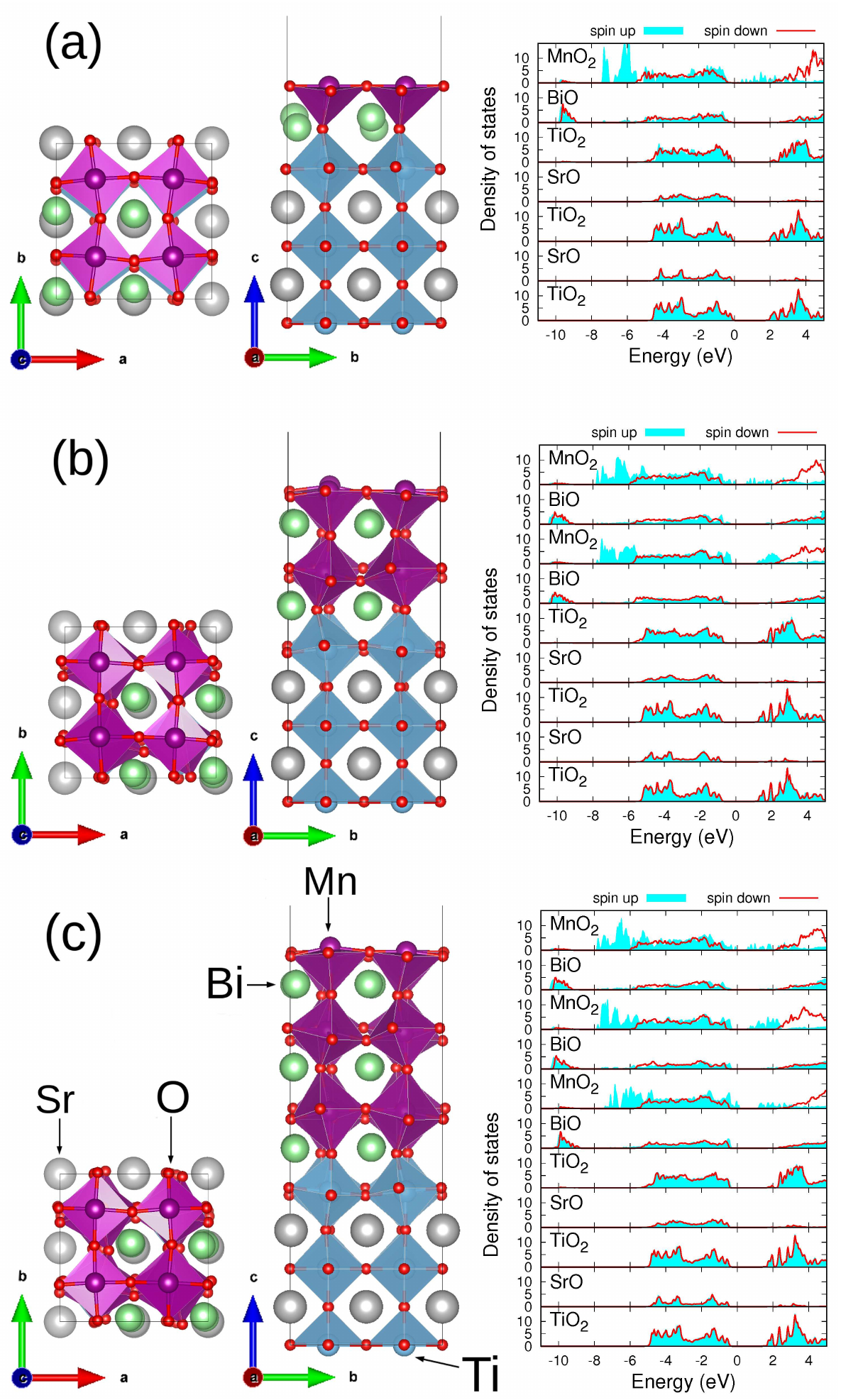}
\caption{
\label{fig:thinfilms}
Atomic structures and layer-projected density of states of 
(a) monolayer, (b) bilayer and (c) trilayer
\BMO{} thin films on \STO{} substrates.
}
\end{center}
\end{figure}

The aim of this study is to address the difficult issue of ferroelectricity in \BMO{} epitaxial thin films 
and to unveil the interplay among structure, electronic and magnetic structure.
To accomplish this goal, we first establish a DFT-based theoretical approach including both Hubbard Coulomb and exchange corrections 
(see Supplementary Material section I.A) that proves capable of providing a satisfactory description of the physical properties of bulk \BMO.
The Hubbard exchange term, overlooked in previous theoretical studies, 
plays a crucial role in describing noncollinear magnets \cite{PhysRevB.82.220402}, 
multiband metals \cite{PhysRevB.83.205112,PhysRevLett.107.256401}, 
Fe-based superconductors \cite{PhysicaC.469.908}, and manganites \cite{PhysRevB.92.085151}.
This is the first key that differentiates this work from previous theoretical studies.
The second key is including explicitly the substrate instead of simulating the epitaxial condition merely by fixing the lateral size of the unit box.
It is reported \cite{APL.93.062902} that \BMO{} thin films lose the epitaxial constraint relation with respect to substrates when the thickness is 
larger than $ \sim 10 \mathrm{nm}$.
In ultrathin films, interface effects not seen in bulk form can be significant, 
such as the emergence of novel physical properties exhibited by perovskite  superlattices and heterointerfaces 
 \cite{Science.305.646,Nature.427.423,MRS.Bulletin.31.28,Science.327.1607}.

\begin{figure}[b]
\begin{center}
\includegraphics[width=\linewidth]{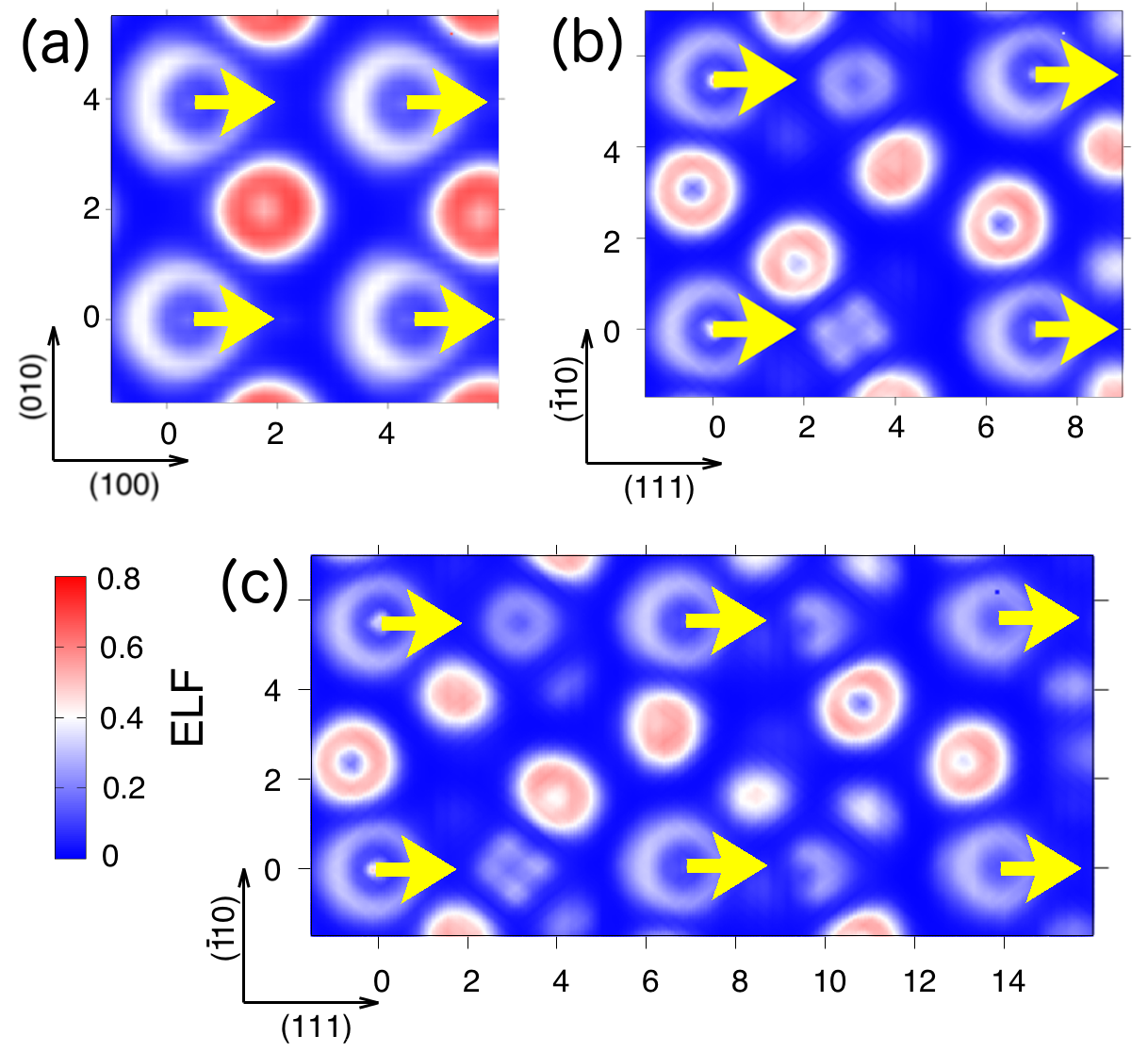}
\caption{
\label{fig:ELF} 
ELF (electron localization function)  of (a) monolayer, (b) bilayer, and (c) trilayer \BMO{} thin films.
The ELF is plotted on a plane crossing the ions. In (a) the plane is parallel to the (100) and (010) directions.
In (b) and (c) the plane is parallel to the (111) and $(\bar{1}10)$ directions.
Bi ions at the surface appear at the right hand side of (b) and (c).
The off-center displacements of Bi ions are denoted by arrows.
The ELF around Bi ions loss spherical symmetry and Bi-$6s^2$ lone pairs are localized in the direction opposite to Bi off-center displacements. 
Other visible features in the ELF plot are contributed by oxygen. 
}
\end{center}
\end{figure}

{\it Computational method.}
Calculations were carried out using the DFT with Hubbard Coulomb and exchange corrections (DFT+U+J) method. The two parameters $U$ and $J$ of this method were determined to reproduce the experimental 
properties of bulk \BMO{}. See Section I of Supplementary Material for more details.

For substrate material we chose \STO, 
which is widely used in experiments \cite{APL.84.91-2004,EPL.74.348-2006,APL.93.062902,PhysRevLett.111.087204,APL.93.062902}.
The ultrathin \BMO{} films are simulated using \BMO{}/\STO{}/\BMO{} slabs 
in which five atomic-layer thick, $\mathrm{TiO_2}$-terminated \STO{} serves as the substrate.
A $2 \times 2$ pseudo-cubic supercell along the in-plane directions is used and the in-plane lattice constant in the central region is fixed 
at the \STO{} substrate bulk value ($3.905 \Ang$).
The epitaxial conditions imposed by the substrate are maintained during structural relaxation.
The slab model is symmetric in the $z$-direction to maximally illuminate dipole-dipole interactions.

{\it Results and discussion.}
Calculated magnetic and vibrational properties of bulk \BMO{} are in good agreement with experimental data.
The detailed data are included in Section II of the Supplementary Material.
The bulk is not ferroelectric, and our analysis shows a cancelation of dipole moments due to 
disordered Bi ion displacements, unlike in \BFO{} in which Bi displacements are aligned.
The reason lies in the co-existence of stereochemically active Bi-$6s^2$ lone pairs 
and Jahn-Teller distortions around Mn ions.
From this, we move on to examine thin film systems.

The atomic structures of the \BMO{}-substrate interfaces are obtained for monolayer, bilayer, and trilayer \BMO{} thin films on \STO{} 
(Fig.~\ref{fig:thinfilms}).   
Structural results confirm the possible existence of ferroelectric order.
We first examine the off-center displacements of Bi ions, which are the main driver necessary for ferroelectricity in \BMO.
In a nutshell, Bi ions occupy the A-site of perovskite structures and are surrounded by twelve oxygens.
Bi ions tend to exhibit off-center displacements due to their $6s^2$ lone pairs \cite{ChemMater.13.2892}.
Among the four inequivalent Bi ions of a \BMO{} monolayer, two of them move along the pseudo-cubic (010) direction by 
$0.6 \Ang$, and the other two along the (011) direction by $0.7 \Ang$.
For thicker \BMO{} films, Bi ions below the surface layer are shifted along the (111) direction by about $0.5 \Ang$, 
while the displacement direction is between (111) and (001) for Bi ions at the surface layer.
The most important observation here is that the displacements of Bi ions are roughly parallel to each other, 
which leads to a polar structure of \BMO{} thin films. This is a key finding of this work.
The stereochemically active Bi-$6s^2$ lone pairs are illustrated by the electron localization function (ELF) \cite{Nature.371.683}.
ELF measures the extent of electron pair probability (see Supplementary Material Section I.C.), and was employed for illustrating lone pairs.
The ELF of \BMO{} thin films is shown in Fig.~\ref{fig:ELF}, in which the off-center displacements of Bi ions are denoted by arrows.
The ELF around each Bi ion shows a loss of spherical symmetry, 
indicating that the spatial distribution of $6s^2$ lone pairs is accumulated in the  direction opposite the ion displacement.

Band structure and density of states analysis shows that energy gaps for mono-, bi-, and 
trilayer systems are $0.20$, $0.40$, and $0.20 \eV $ respectively.
Our calculations also show that these three \BMO{} systems are ferromagnetic, and we thus conclude that 
 \BMO{} thin films on \STO{} are multiferroic.
Detailed analysis of magnetic moments will be given later.

In order to understand the mechanism for the polar structure of \BMO{} films, 
we compare off-center displacements of Bi ions with those in bulk \BMO{} and bulk \BFO
(detailed data are in Section IIA Supplementary Material).
In bulk C2/c \BMO, Bi displacements are close to the pseudo-cubic $(011)$ and $(0\bar{1} \bar{1})$ directions, 
and they cancel with each other, for a  nonpolar structure.
In \BMO{} thin films thicker than monolayer, \BMO{} layers below the surface layer are the most bulk-like, 
but the Bi displacements are along the (111) direction and are parallel to each other.
Interface with the substrate not only alters the direction of Bi off-center displacement, 
but also shifts the antiparallel pattern into a parallel one.
Note that the (111) direction is the direction of Bi displacement in another multiferroic material, \BFO.
Comparison between bulk \BMO{} and \BFO{} gives clearer hints on the differences between bulk and thin films of \BMO.
\BFO{} exhibits a more ordered structure and its unit cell contains only two pseudo-cubic cells (eight cells for bulk \BMO{}).
Chemically the difference between \BMO{} and \BFO{} is the much weaker Jahn-Teller distortion in \BFO.
The strong disorder in \BMO{} results from the competition between Jahn-Teller distortion around Mn ions 
and the off-center displacement of Bi ions.
It is the Jahn-Teller distortion that drives the Bi ion displacement in (011)  and antiparallel directions 
(FIG.~2 of Supplementary Material).
In \BMO{} thin films the Bi ion displacements resemble those in \BFO{} because the interaction with the substrate 
reduces significantly the disorder in the Jahn-Teller distortions.
Besides \BMO/\STO{} interfaces, surfaces of \BMO{} thin films also affect Bi ions displacements.
While the Bi ions below the surface layer displace almost along the (111) direction, 
those in the surface layer displace along a direction between (111) and (001) directions. 

Besides the spontaneous polarization mentioned above, 
a relatively low energy barrier for polarization rotation is another indicator of ferroelectricity.
The energy barrier measures the external electric field strength required to rotate the direction of the polarization.
The calculated energy barrier for a $90^\circ$ rotation of the polarization is about $80 \meV$ per \BMO{} formula unit (f.u.),
a value that is 3 to 4 times larger than the $20 \meV$ for   $\mathrm{BaTiO_3}$ \cite{PhysRevB.42.6416}, 
$30 \meV$ for for $\mathrm{PbTiO_3}$ \cite{PhysRevB.55.6161}, 
and $25 \meV$ in hexagonal $R\mathrm{MnO_3}$ \cite{NatComm.4.1540}.

Next we turn to the valence states and magnetic moments of $\Mn$ ions.
Charge disproportionation, the instability of 
$\Mn^{3+}$ valence states  ($m =3.7 \,\mu_B$) to form  $\Mn^{2+}$ ($m=4.5  \, \mu_B$) and 
$\Mn^{4+}$  ($m=3.0  \, \mu_B$), occurs in \BMO{} thin films except the monolayer case.
The valence states $\Mn^{2+}$ and $\Mn^{4+}$ always appear in pairs, 
and are not formed by charge transfer between \BMO{} thin films and the substrate.
The density of states projected onto the \STO{} substrate (not shown) shows that it remains insulating, 
so little charge transfer happens between the \STO{} substrate and \BMO{} thin films.
For a given pair, the $\Mn^{2+}$ ion resides in the surface layer and the $\Mn^{4+}$ ion is located in the layer below it; 
they occupy the diagonal sites of one vertical face of a pseudo-cubic cell (see Supplementary Material Section III.B).
The Mn-O bonds around the $\Mn^{2+}$ ion have different lengths: one is $2.45 \Ang$ and the other four are $2.05 \Ang$.
The six $\Mn^{4+}$-O bonds have similar bond lengths of $1.90$--$2.00 \Ang$.
As a reference, for $\Mn^{3+}$ ions, two of $\Mn^{3+}$-O bonds are between $2.15$ and $2.25 \Ang$, 
and the other four are between $1.90$ and $2.00 \Ang$.
The volume occupied by Mn ions in different valence states is $\Mn^{2+} > \Mn^{3+} > \Mn^{4+}$.
The charge disproportionation thus introduces a local strain to the lattice, 
which explains the occurrence of $\Mn^{2+}$-$\Mn^{4+}$ pairs only at the surface.

The occurrence of $\Mn^{2+}$ at the surface was observed in other perovskite manganites 
\cite{PhysRevB.55.11511,PhysRevB.73.104402}.
Two possible mechanisms were proposed in the literature; 
one was attributed to the instability of $\Mn^{3+}$ via $ 2\,\Mn^{3+} \to \Mn^{2+} + \Mn^{4+}$ \cite{PhysRevB.55.11511}, 
while ref.~\cite{PhysRevB.73.104402} proposed that $\Mn^{2+}$ ions result from chemical reduction 
by electron-doping or removing surface oxygens.
The second mechanism may apply for the \BMO{} thin films, since Mn ions the at surface 
are bonded to five oxygens instead of six in the bulk.
However this fails to explain the absence of $\Mn^{2+}$ in monolayer \BMO{} thin film.
Electron-doping can be excluded because \BMO{} thin films with $\Mn^{2+}$ ions are still insulating.
Nevertheless, the observation that $\Mn^{2+}$ and $\Mn^{4+}$ always occur in pairs support the $\Mn^{3+}$ instability mechanism.
The local strain introduced by spontaneous charge disproportionation sheds light 
on understanding the difficulty during the experimental synthesis of \BMO{} thin films.

\begin{figure}[t]
\begin{center}
\includegraphics[width=0.9\linewidth]{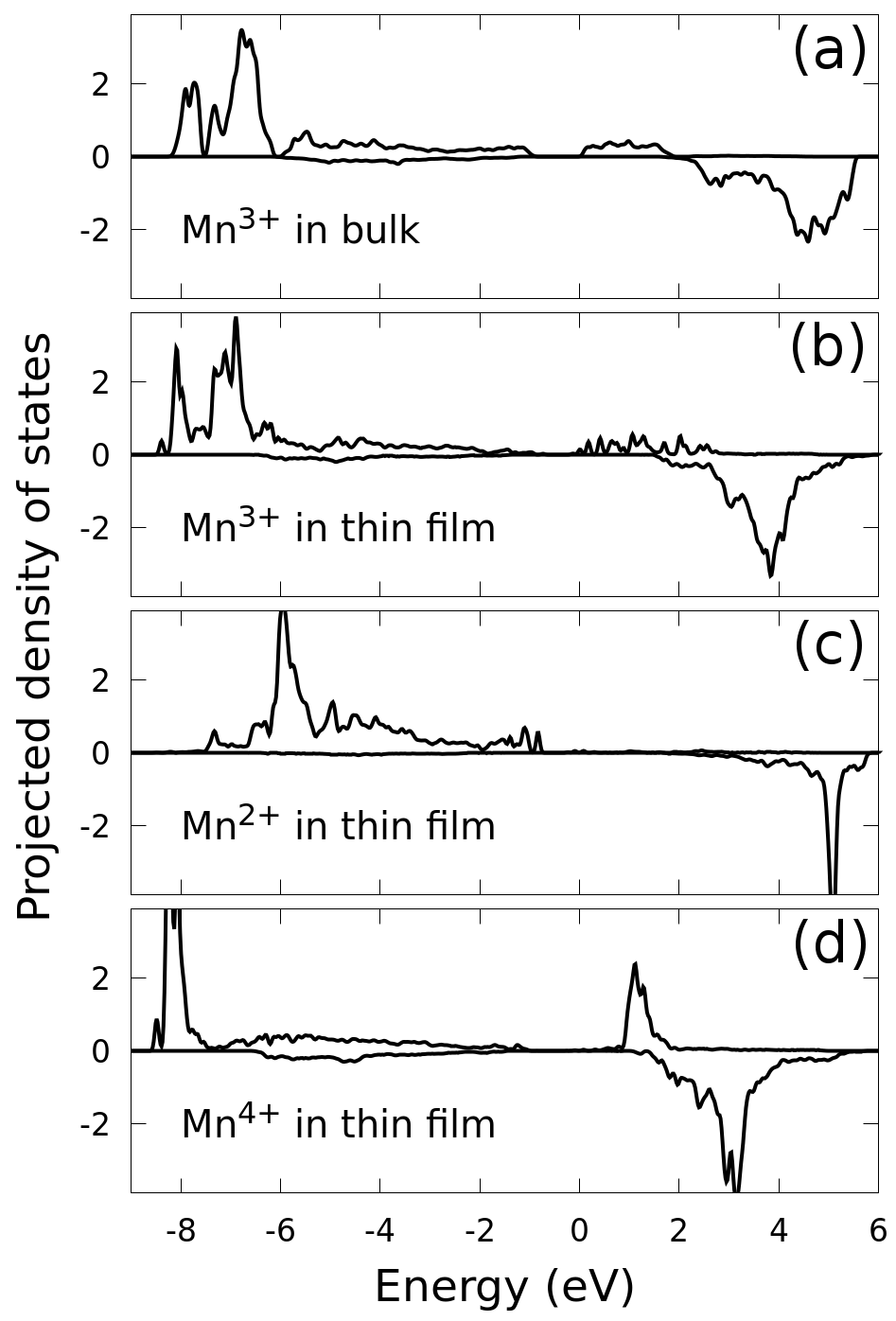}
\caption{
\label{fig:Mn-pdos}
Density of states projected onto Mn ions:
(a) $\Mn^{3+}$ ions in bulk C2/c \BMO{} and
(b) $\Mn^{3+}$,
(c) $\Mn^{2+}$,
(d) $\Mn^{4+}$ ions in bilayer \BMO{} thin film.
}
\end{center}
\end{figure}

The charge disproportionation of Mn ions is manifested in their projected density of states (PDOS).
In Fig.~\ref{fig:Mn-pdos}, we compared the PDOS on {\it d}-orbitals of $\Mn^{3+}$ ions in bulk C2/c \BMO{} 
and $\Mn^{3+}$, $\Mn^{2+}$, and $\Mn^{4+}$ ions in bilayer \BMO{} thin films.
The {\it d}-shells of $\Mn^{2+}$, $\Mn^{3+}$ and $\Mn^{4+}$ are $d^5$, $d^4$, and $d^3$ respectively.
Due to Hund's rule, all the {\it d}-electrons occupy the same spin channel.
The crystal field around Mn ions splits the 10 {\it d}-orbitals into triply degenerate $t_{2g}$ and doubly degenerate $e_g$ orbitals.
The two $e_{g}$ orbitals further split due to Jahn-Teller distortions.
The four $d$-electrons of $\Mn^{3+}$ ions in bulk \BMO{} [Fig.~\ref{fig:Mn-pdos}(a)] are mainly located 
between $-8 \eV$ and $-6 \eV$.
The PDOS between $-6 \eV$ and $-1 \eV$ is contributed by hybridizations with O-{\it p} orbitals, 
and the PDOS in the spin-up channel above the energy gap is contributed by $d_{x^2-y^2}$ orbitals.
Ions $\Mn^{3+}$ in thin films [Fig.~\ref{fig:Mn-pdos}(b)] exhibit a PDOS very similar to that in bulk \BMO, 
as expected, but on the other hand  $\Mn^{2+}$ and $\Mn^{4+}$ ions show a very different PDOS.
All five of the {\it d}-orbitals of the $\Mn^{2+}$ ion are occupied.
The peak in the PDOS in the $\Mn^{4+}$ spin-up channel (unoccupied) is contributed by $t_{2g}$ orbitals.

{\it Summary and outlook.}
First-principles simulations of \BMO{} thin films on \STO{} substrates confirm the presence of 
ferroelectric order in monolayer, bilayer, and trilayer \BMO{} thin films.
Our calculations show that these thin films are indeed multi-ferroic, that is, they exhibit both ferroelectric and magnetic behavior.
The ferroelectricity originates  from the ordered displacement of Bi ions from the the tetrahedron centers, 
a consequence of the substrate constraint.
Magnetic moments, density of states, and orbital analysis are presented to support our conclusions.
We have resolved a long-standing physical problem of general interest and our observations point to  
a general principle in designing multiferroic materials by ordering dipole moments using substrate constraints. 
Hopefully this work will stimulate future experiments in search of multi-functional materials systems based on \BMO{}. 

{\it Acknowledgments.}
This work was supported by the US Department of Energy
(DOE), Office of Basic Energy Sciences (BES), under Contract
No. DE-FG02-02ER45995.
Computations were done using the utilities of the
National Energy Research Scientific Computing Center (NERSC).



%


\end{document}